\documentclass[final]{ias2}

\usepackage{graphicx} 
\usepackage{multirow}
\usepackage{array} 

\usepackage{hyperref} 

\begin{document}

\markboth{Effective interaction . . . . }{D. N. Basu}

\title{Effective interaction:From nuclear reactions to neutron stars}

\author[x]{D. N. Basu}
\email{dnb@vecc.gov.in}
\address[x]{Variable  Energy  Cyclotron  Centre, 1/AF Bidhan Nagar, Kolkata 700 064, India}

\begin{abstract}
    An equation of state (EoS) for symmetric nuclear matter is constructed using the density dependent M3Y effective interaction and extended for isospin asymmetric nuclear matter. Theoretically obtained values of symmetric nuclear matter incompressibility, isobaric incompressibility, symmetry energy and its slope agree well with experimentally extracted values. Folded microscopic potentials using this effective interaction, whose density dependence is determined from nuclear matter calculations, provide excellent descriptions for proton, alpha and cluster radioactivities, elastic and inelastic scattering. The nuclear deformation parameters extracted from inelastic scattering of protons agree well with other available results. The high density behavior of symmetric and asymmetric nuclear matter satisfies the constraints from the observed flow data of heavy-ion collisions. The neutron star properties studied using $\beta$-equilibrated neutron star matter obtained from this effective interaction reconcile with the recent observations of the massive compact stars.
\end{abstract}

 
\maketitle

\vspace{-0.2cm}
\section{Introduction}

    The measurements of nuclear masses, densities and collective excitations have allowed to resolve some of the basic features of equation of state (EoS) of nuclear matter. However, the symmetry properties of EoS due to differing neutron and proton numbers remain more elusive to date and study of the isospin dependent properties of asymmetric nuclear matter and the density dependence of the nuclear symmetry energy (NSE) have become the prime objective \cite{Li08,St05}. Consequently, the ultimate goal of such study is to extract information on the isospin dependence of in-medium nuclear effective interactions as well as the EoS of isospin asymmetric nuclear matter, particularly its isospin-dependent term or the density dependence of the NSE. This knowledge is important for understanding the structure of radioactive nuclei, the reaction dynamics induced by rare isotopes, the liquid-gas phase transition in asymmetric nuclear matter and many critical issues in astrophysics \cite{Li08,St05,Da02}. 

    In this work, based on the theoretical description of nuclear matter using the density dependent M3Y-Reid-Elliott effective interaction \cite{Be77,Sa79} (DDM3Y), we carry out a systematic study of the symmetric nuclear matter (SNM) and isospin-dependent bulk properties of asymmetric nuclear matter. In particular, we study the density dependence of the NSE and extract the slope $L$ and the curvature $K_{sym}$ parameters of the NSE and the isospin dependent part $K_\tau$ of the isobaric incompressibility. 
    
    The lifetimes of radioactive decays are calculated theoretically within the improved WKB approximation \cite{ke35} using microscopic proton, $\alpha$ and nucleus-nucleus interaction potentials. These nuclear potentials have been obtained by folding the densities of the emitted and the daughter nuclei with the M3Y effective interaction, whose density dependence is determined from nuclear matter calculations. These calculations provide reasonable estimates of half-lives for the observed proton \cite{BCS08}, $\alpha$ \cite{Ba03,CSB06,prc07,scb07} and cluster \cite{Ro09} radioactivities. These folding model potentials provide excellent descriptions for elastic and inelastic scattering and the nuclear deformation parameters extracted from inelastic scattering of protons \cite{Gu05,Gu06} agree well with other available results. 

    We present a systematic study of the properties of pure hadronic and hybrid compact stars. The nuclear EoS for $\beta$-equilibrated neutron star (NS) matter obtained using density dependent effective nucleon-nucleon interaction satisfies the constraints from the observed flow data from heavy-ion collisions. The energy density of quark matter is lower than that of this nuclear EoS at higher densities implying possibility of transition to quark matter inside the core. We solve the Einstein's equations for rotating stars using pure nuclear matter and quark core. The $\beta$- equilibrated neutron star matter with a thin crust is able to describe highly massive compact stars \cite{De10} but find that the nuclear to quark matter deconfinement transition inside neutron stars causes reduction in their masses. 
    
\vspace{-0.2cm}    
\section{Effective interaction $\&$ its density dependence from nuclear matter calculations}

    The nuclear matter EoS is calculated using isoscalar and isovector \cite{Sa83} components of M3Y effective nucleon-nucleon interaction along with density dependence. The density dependence of the effective interaction, DDM3Y, is completely determined from nuclear matter calculations. The equilibrium density of the nuclear matter is determined by minimizing the energy per nucleon. The energy variation of the zero range potential is treated accurately by allowing it to vary freely with the kinetic energy part $\epsilon^{kin}$ of the energy per nucleon $\epsilon$ over the entire range of $\epsilon$. This is not only more plausible, but also yields excellent result for the incompressibility $K_\infty$ of SNM which does not suffer from the superluminosity problem. 
        
    In a Fermi gas model of interacting neutrons and protons, with isospin asymmetry $X$=$\frac{\rho_n - \rho_p} {\rho_n + \rho_p}$, $\rho$=$\rho_n$+$\rho_p$ where $\rho_n$, $\rho_p$ and $\rho$ are the neutron, proton and nucleonic densities respectively, energy per nucleon for isospin asymmetric nuclear matter is given by \cite{BCS08}

\vspace{-0.4cm}
\begin{equation}
 \epsilon(\rho,X) = [\frac{3\hbar^2k_F^2}{10m}] F(X) + (\frac{\rho J_v C}{2}) (1 - \beta\rho^n)  
\label{seqn1}
\end{equation}
\vspace{-0.4cm}

\noindent
where $k_F$=$(1.5\pi^2\rho)^{\frac{1}{3}}$ which equals Fermi momentum in case of SNM, the kinetic energy per nucleon $\epsilon^{kin}$=$[\frac{3\hbar^2k_F^2}{10m}] F(X)$ with $F(X)$=$[\frac{(1+X)^{5/3} + (1-X)^{5/3}}{2}]$ and $J_v$=$J_{v00} + X^2 J_{v01}$, $J_{v00}$ and $J_{v01}$ represent the volume integrals of the isoscalar and the isovector parts of the M3Y interaction. The isoscalar $t_{00}^{M3Y}$ and the isovector $t_{01}^{M3Y}$ components of M3Y interaction potential are given by $t_{00}^{M3Y}(s, \epsilon)$=7999$\frac{\exp( - 4s)}{4s}-2134\frac{\exp( - 2.5s)}{2.5s}$+$J_{00}$(1$-\alpha\epsilon$)$\delta(s)$ and $t_{01}^{M3Y}(s, \epsilon)$=$-4886\frac{\exp( - 4s)}{4s}$+$1176\frac{\exp( - 2.5s)}{2.5s}$+$J_{01}$(1$-\alpha\epsilon$)$\delta(s)$ with $J_{00}$=$-$276 MeVfm$^3$, $J_{01}$=228 MeVfm$^3$, $\alpha=0.005$MeV$^{-1}$. The DDM3Y effective NN interaction is given by $v_{0i}(s,\rho, \epsilon) = t_{0i}^{M3Y}(s, \epsilon) g(\rho)$ where the density dependence $g(\rho) = C (1 - \beta \rho^n)$ with $C$ and $\beta$ being the constants of density dependence. 
        
    The Eq.(1) along with the saturation condition $\frac{\partial\epsilon}{\partial\rho} = 0$ at $X=0$, $\rho = \rho_{0}$, $\epsilon = \epsilon_{0}$ can be solved simultaneously for fixed values of the saturation energy per nucleon $\epsilon_0$ and the saturation density $\rho_{0}$ of the cold SNM to obtain the values of $\beta$ and $C$. The constants of density dependence $\beta$ and $C$, thus obtained, are given by 
\vspace{-0.2cm}
\begin{equation}
 \beta = \frac{[(1-p)+(q-\frac{3q}{p})]\rho_{0}^{-n}}{[(3n+1)-(n+1)p+(q-\frac{3q}{p})]},~p = \frac{[10m\epsilon_0]}{[\hbar^2k_{F_0}^2]},~q=\frac{2\alpha\epsilon_0J_{00}}{J^0_{v00}}
\label{seqn3}
\end{equation} 
\noindent
where $J^0_{v00} = J_{v00}(\epsilon^{kin}_0)$ which means $J_{v00}$ at $\epsilon^{kin}=\epsilon^{kin}_0$, the kinetic energy part of the saturation energy per nucleon of SNM,  $k_{F_0} = [1.5\pi^2\rho_0]^{1/3}$ and 
\vspace{-0.2cm}
\begin{equation}
 C = -\frac{[2\hbar^2k_{F_0}^2] }{ 5mJ^0_{v00} \rho_0 [1 - (n+1)\beta\rho_0^n -\frac{q\hbar^2k_{F_0}^2 (1-\beta\rho_0^n)}{10m\epsilon_0}]},
\label{seqn5}
\end{equation} 
\noindent
respectively. It is quite obvious that the constants of density dependence $C$ and $\beta$ obtained by this method depend on the saturation energy per nucleon $\epsilon_0$, the saturation density $\rho_{0}$, the index $n$ of the density dependent part and on the strengths of the M3Y interaction through the volume integral $J^0_{v00}$. 

    The calculations are performed using the values of the saturation density $\rho_0$=0.1533 fm$^{-3}$ \cite{Sa89} and the saturation energy per nucleon $\epsilon_0$=$-$15.26 MeV \cite{CB06} for the SNM obtained from the co-efficient of the volume term of Bethe-Weizs\"acker mass formula which is evaluated by fitting the recent experimental and estimated atomic mass excesses from Audi-Wapstra-Thibault atomic mass table \cite{Au03} by minimizing the mean square deviation incorporating correction for the electronic binding energy \cite{Lu03}. Using the usual values of $\alpha$=0.005 MeV$^{-1}$ for the parameter of energy dependence of the zero range potential and $n$=2/3 and $-15.26\pm$0.52 MeV for the saturation energy per nucleon (which is the volume energy coefficient $a_v$ and covers, more or less, the entire range of values obtained for $a_v$ \cite{Ro06}), values obtained for the constants of density dependence are $C$=2.2497$\pm$0.0420, $\beta$=1.5934$\pm$0.0085 fm$^2$ and that for SNM incompressibility is $K_\infty$=274.7$\pm$7.4 MeV.  

\vspace{-0.2cm}      
\section{The equation of state}

\vspace{-0.2cm}
\subsection{Symmetric and asymmetric nuclear matter}

    The EoSs of the symmetric and the asymmetric nuclear matter describing energy per nucleon as a function of nucleonic density can be obtained by setting $X=0$ and non-zero, respectively, in Eq.(1). The incompressibility or the compression modulus of the SNM, which is a measure of the curvature of an EoS at saturation density and defined as $k_F^2\frac{\partial^2\epsilon}{\partial{k_F^2}} \mid_{k_F=k_{F_0}}$, measures the stiffness of an EoS and obtained theoretically using Eq.(1) for $X$=0. The incompressibilities for isospin asymmetric nuclear matter are evaluated at saturation densities $\rho_s$ with the condition $\frac{\partial\epsilon}{\partial\rho}$=0 which corresponds to vanishing pressure. The incompressibility $K_s$ for isospin asymmetric nuclear matter is therefore expressed as 
\vspace{-0.4cm}
\begin{eqnarray}
 &&K_s = -\frac{3\hbar^2k_{F_s}^2}{5m} F(X) - \frac{9 J^s_v C n(n+1) \beta\rho_s^{n+1}}{2} - 9\alpha J C [1-(n+1)\beta\rho_s^n] \nonumber \\
 &&\times [\frac{\rho_s\hbar^2k_{F_s}^2}{5m}] F(X) + [\frac{3\rho_s\alpha J C (1-\beta\rho_s^n)\hbar^2k_{F_s}^2}{10m}] F(X). 
\label{seqn6}
\end{eqnarray} 
\noindent
Here $k_{F_s}$ means that the $k_F$ is evaluated at the saturation density $\rho_s$. $J^s_v$=$J^s_{v00}$+$X^2 J^s_{v01}$ is $J_v$ at $\epsilon^{kin}$=$\epsilon^{kin}_s$ which is the kinetic energy part of the saturation energy per nucleon $\epsilon_s$ and $J$=$J_{00}$+$X^2$$J_{01}$.

    In Table-1 incompressibility of isospin asymmetric nuclear matter  $K_s$ as a function of the isospin asymmetry parameter $X$, is provided. The magnitude of the incompressibility $K_s$ decreases with the isospin asymmetry $X$ due to lowering of the saturation densities $\rho_s$ with $X$ as well as decrease in the EoS curvature. At high isospin asymmetry $X$, the isospin asymmetric nuclear matter does not have a minimum signifying that it can never be bound by itself due to nuclear interaction. However, the $\beta$ equilibrated nuclear matter which is highly neutron rich asymmetric nuclear matter exists in the core of neutron stars since its E/A is lower than that of SNM at high densities and is unbound by the nuclear force but can be bound due to high gravitational field realizable inside neutron stars.      

\begin{table}
\centering
\caption{Incompressibility of isospin asymmetric nuclear matter using $\rho_0$=0.1533 fm$^{-3}$, $\epsilon_0$=-15.26 MeV, $n$=2/3 and $\alpha$=0.005 MeV$^{-1}$.}
\begin{tabular}{ccc}
\hline
\hline
$X$&$\rho_s$& $K_s$      \\
\hline
 & fm$^{-3}$ &MeV    \\ 
\hline
 0.0&0.1533&274.7 \\ 
 0.1&0.1525&270.4 \\ 
 0.2&0.1500&257.7 \\ 
 0.3&0.1457&236.6 \\ 
 0.4&0.1392&207.6 \\ 
 0.5&0.1300&171.2 \\  \hline
\hline
\label{table1}
\end{tabular} 
\end{table}
     
    It is interesting to note that the RMF-NL3 incompressibility for SNM is 271.76 MeV \cite{La99} which is about the same as 274.7$\pm$7.4 MeV obtained from present calculation. The recent acceptable value \cite{Vr03,Sh09} of SNM incompressibility lies in the range of 250-270 MeV and calculated value of 274.7$\pm$7.4 MeV is a good theoretical result and is only slightly too high. Although, parameters of the density dependence of DDM3Y interaction have been tuned to reproduce $\rho_0$ and $\epsilon_0$ which are obtained from finite nuclei, the agreement of the present EoS with the experimental flow data \cite{Da02}, where the high density behaviour looks phenomenologically confirmed, justifies its extrapolation to high density.

\vspace{-0.2cm} 
\subsection{Incompressibility, isobaric incompressibility, symmetry energy and its slope}

    EoS of isospin asymmetric nuclear matter, given by Eq.(1) can be expanded in general as
\vspace{-0.28cm}
\begin{equation}
  \epsilon(\rho,X) =  \epsilon(\rho,0) + E_{sym}(\rho) X^2 + O ( X^4)
\label{seqn7}
\end{equation}
\vspace{-0.4cm}

\noindent
and $E_{sym}(\rho)= \frac{1}{2} \frac{\partial^2\epsilon(\rho,X)}{\partial{X^2}} \mid_{X=0}$ is termed as the NSE. The absence of odd-order terms in $X$ in Eq.(5) is due to the exchange symmetry between protons and neutrons in nuclear matter when one neglects the Coulomb interaction and assumes the charge symmetry of nuclear forces. The higher-order terms in $X$ are negligible and to a good approximation, the density-dependent NSE $E_{sym}(\rho)$ can be extracted using following equation \cite{Kl06}
\vspace{-0.2cm}
\begin{equation}
 E_{sym}(\rho)=\epsilon(\rho,1) -\epsilon(\rho,0)
\label{seqn8}
\end{equation}
\vspace{-0.4cm}

\noindent
which can be obtained using Eq.(1) and represents a penalty levied on the system as it departs from the symmetric limit of equal number of protons and neutrons and can be defined as the energy required per nucleon to change SNM to pure neutron matter (PNM).  

    The volume symmetry energy coefficient $S_v$ extracted from nuclear masses provides a constraint on the NSE at nuclear density $E_{sym}(\rho_0)$. The value of $S_v$=30.048 $\pm$0.004 MeV extracted \cite{Mu07} from the measured atomic mass excesses of 2228 nuclei is reasonably close to the theoretical estimate of the value of NSE at saturation density $E_{sym}(\rho_0)$=30.71$\pm$0.26 MeV obtained from the present calculations using DDM3Y interaction. If one uses the alternative definition of $E_{sym}(\rho)= \frac{1}{2} \frac{\partial^2\epsilon(\rho,X)}{\partial{X^2}} \mid_{X=0}$, the value of NSE at saturation density remains almost the same which is 30.03$\pm$0.26 MeV. Empirically the value of $E_{sym}(\rho_0)\approx$ 30 MeV \cite{Da03,St05,Po03} seems well established. Theoretically different parametrizations of the relativistic mean-field (RMF) models, which fit observables for isospin symmetric nuclei well, lead to a relatively wide range of predictions of 24-40 MeV for $E_{sym}(\rho_0)$. The present result of 30.71$\pm$0.26 MeV is close to that using Skyrme interaction SkMP (29.9 MeV) \cite{Be89} and Av18+$\delta v$+UIX$^*$ variational calculation (30.1 MeV) \cite{Ak98}.  
 
    Around the nuclear matter saturation density $\rho_0$ the NSE $E_{sym}(\rho)$ can be expanded to second order in density as 
\vspace{-0.4cm}
\begin{equation}
 E_{sym}(\rho)= E_{sym}(\rho_0) + \frac{L}{3} {\Big (}\frac{\rho - \rho_0}{\rho_0}{\Big )}+ \frac{K_{sym}}{18}{\Big (}\frac{\rho - \rho_0}{\rho_0}{\Big )}^2
\label{seqn9}
\end{equation}
\vspace{-0.7cm}

\noindent
where $L$ and $K_{sym}$ represents the slope and curvature parameters of NSE at $\rho_0$ and hence $L= 3\rho_0 \frac{\partial E_{sym}(\rho)}{\partial\rho} \mid_{\rho=\rho_0}$ and $K_{sym}= 9\rho_0^2 \frac{\partial^2 E_{sym}(\rho)}{\partial {\rho^2}} \mid_{\rho=\rho_0}$. The $L$ and $K_{sym}$ characterize the density dependence of the NSE around normal nuclear matter density and thus carry important information on the properties of NSE at both high and low densities. In particular, the slope parameter $L$ has been found to correlate linearly with the neutron-skin thickness of heavy nuclei and can be determined from the measured thickness of neutron skin of such nuclei \cite{Ce09}. The isobaric incompressibility for infinite nuclear matter can be expanded in the power series of isospin asymmetry $X$ as $K_\infty (X) = K_\infty + K_\tau X^2 + K_4 X^4 + O(X^6)$. The magnitude of the higher-order $K_4$ parameter is generally quite small compared to $K_\tau$ \cite{Ch09}. The latter characterizes the isospin dependence of the incompressibility at saturation density and can be expressed as $K_\tau=K_{sym}-6L-\frac{Q_0}{K_\infty}L = K_{asy}-\frac{Q_0}{K_\infty}L$ where $Q_0$ is the third-order derivative parameter of SNM at $\rho_0$ given by $Q_0 = 27\rho_0^3 \frac{\partial^3 \epsilon(\rho,0)}{\partial {\rho^3}} \mid_{\rho=\rho_0}$. In Table-2, the values of $K_\infty$, $E_{sym}(\rho_0)$, $L$, $K_{sym}$ and $K_\tau$ are listed and compared with the corresponding quantities obtained with relativistic mean field (RMF) models \cite{Pi09}. In Fig.-1, $K_\tau$ is plotted against $K_\infty$ for the present calculation using DDM3Y interaction and compared with the predictions of FSUGold, NL3, Hybrid \cite{Pi09}, SkI3, SkI4, SLy4, SkM, SkM*, NLSH, TM1, TM2, DDME1 and DDME2 as given in Table-1 of Ref.\cite{Sa07}. The dotted rectangular region encompasses the recent values of $K_\infty$=250-270 MeV \cite{Sh09} and $K_\tau$=$-$370$\pm$120 MeV \cite{Ch09}. Although both DDM3Y and SkI3 are within the above region, unlike DDM3Y the $L$ value for SkI3 is 100.49 MeV which is much above the acceptable limit of 45-75 MeV \cite{Wa09} whereas DDME2 which gives $L$=51 MeV is reasonably close to the rectangular region. Present NSE is `super-soft' because it increases initially with nucleonic density up to about two times the normal nuclear density and then decreases monotonically (hence `soft') and becomes negative (hence `super-soft') at higher densities (about 4.7 times the normal nuclear density) \cite{BCS08,CBS09} and is consistent with recent evidence for a soft NSE at suprasaturation densities \cite{Zh09} and with the fact that super-soft nuclear symmetry energy preferred by the FOPI/GSI experimental data on $\pi^+/\pi^-$ ratio in relativistic heavy-ion reactions can readily keep neutron stars stable if the non-Newtonian gravity proposed in the grand unification theories is considered \cite{We09}. 
    
\vspace{-0.54cm}
\begin{table*}[hb]
\centering
\caption{Results of present calculations (DDM3Y) for SNM incompressibility $K_\infty$, nuclear symmetry energy $E_{sym}(\rho_0)$, slope $L$ and curvature $K_{sym}$ of nuclear symmetry energy, approximate isospin dependent part $K_{asy}$ and exact part $K_\tau$ of isobaric incompressibility (all in MeV) compared with those obtained with RMF models \cite{Pi09}.}
\begin{tabular}{cccccccc}
\hline
\hline
Model&$K_\infty$&$E_{sym}(\rho_0)$&$L$&$K_{sym}$&$K_{asy}$&$Q_0$&$K_\tau$\\ 
\hline
 This work&274.7&30.71&45.11&-183.7&-454.4&-276.5&-408.97 \\ 
 &$\pm$7.4&$\pm$0.26&$\pm$0.02&$\pm$3.6&$\pm$3.5&$\pm$10.5&$\pm$3.01 \\
 FSUGold&230.0&32.59&60.5&-51.3&-414.3&-523.4&-276.77\\
 NL3&271.5&37.29&118.2&+100.9&-608.3&+204.2&-697.36 \\
 Hybrid&230.0&37.30&118.6&+110.9&-600.7&-71.5&-563.86\\ 
\hline
\hline
\label{table2}
\end{tabular} 
\end{table*}

\vspace{0.0cm}
\begin{figure}[ht]
\begin{center}
\includegraphics[width=0.59\columnwidth]{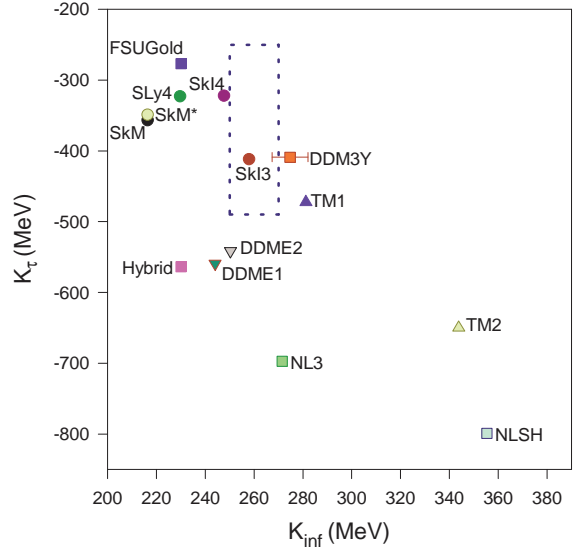}
\caption{$K_\tau$ is plotted against $K_\infty$ ($K_{inf}$) for present calculations using DDM3Y interaction and compared with other predictions \cite{Pi09,Sa07}. The dotted rectangular region encompasses the values of $K_\infty$=250-270 MeV \cite{Sh09} and $K_\tau$=$-370\pm120$ MeV \cite{Ch09}.}
\label{fig:5}
\end{center}
\end{figure}
\vspace{0.0cm}

\vspace{-0.28cm}   
\section{Nuclear Scattering}

\vspace{-0.28cm}
\subsection{Elastic scattering using potentials from folding effective interaction}

    The microscopic proton-nucleus interaction potentials are obtained by single folding the density distribution of the nucleus with the DDM3Y effective interaction as following
\vspace{-0.4cm}
\begin{equation}
V_N(R) = \int \rho(\vec{r}) v_{00}(|\vec{R}-\vec{r}|)d^3\vec{r} \\
\label{seqn15}
\end{equation}
\vspace{-0.52cm}

\noindent
where $\rho(\vec{r})$ is density of the nucleus at $\vec{r}$ and $v_{00}$ is the effective interaction between two nucleons at the sites $\vec{R}$ and $\vec{r}$. The parameters of the density dependence, $C$=2.2497 and $\beta$=1.5934 fm$^2$, used here are obtained from the nuclear matter calculations. The nuclear ground state densities are calculated in the framework of spherical Hartree Fock plus BCS calculations in co-ordinate space using SkM* \cite{BA82} parameterization and used for calculating $V_N(R)$ and form factor. Phenomenological optical potentials have the form $V_{\rm pheno}(r)$=$-V_o~f_o(r)-iW_vf_v(r)$ +$4ia_s W_s\frac{d f_s(r)}{dr}$ +$2(\frac{\hbar}{m_{\pi}c})^2 V_{s.o} (\frac{1}{r}) \frac{d f_{s.o}(r)}{dr} ({\bf L.S})$ +$ V_{\rm coul}$ \\
where $f_x(r)$=$[1+exp(\frac{r-R_x}{a_x})]^{-1}$, $R_{x}$=$r_{x}A^{1/3}$ and $x$=$o,v,s,s.o$. The subscripts $o,v,s,s.o$ denote real, volume imaginary, surface imaginary, spin-orbit respectively and $V_o$, $W_v(W_s)$ and $V_{s.o}$ are the strengths of real, volume (surface) imaginary and spin-orbit potentials respectively. $V_{\rm coul}$ is the Coulomb potential of a uniformly charged sphere of radius 1.20~$A^{1/3}$. In semi-microscopic analysis both the volume real ($V$) and volume imaginary ($W$) parts of the potentials (generated microscopically by folding model) are assumed to have the same shape, i.e. $V_{\rm micro}(r)$=$V$+$iW$=($N_{\rm R}$+$iN_{\rm I}$)$V_N$($r$) where, $N_{\rm R}$ and $N_{\rm I}$ are the renormalization factors for real and imaginary parts respectively~\cite{SA97}. Thus the potentials for elastic scattering analysis include real and volume imaginary terms (folded potentials) and also surface imaginary and spin-orbit terms (best fit phenomenological potentials). Best fits are obtained by minimizing $\chi^2/N$=$\frac{1}{N}\sum_{k = 1}^{\rm N} \left[\frac{\sigma_{th}(\theta_k)-\sigma_{ex}(\theta_k)} {\Delta\sigma_{ex} (\theta_k)}\right]^2$ for each angular distribution, where $\sigma_{th}$, $\sigma_{ex}$ are theoretical and experimental cross sections respectively, at angle $\theta_k$, $\Delta\sigma_{ex}$ is experimental error and N is the number of data points. 

\vspace{-0.2cm}
\subsection{Inelastic scattering and nuclear deformation parameter} 

    The potentials for elastic scattering analysis are subsequently used in the DWBA calculations of inelastic scattering with transferred angular momentum $l$. The calculations are performed using the code DWUCK4~\cite{DWUCK4}. The derivative of the potentials ($\delta \frac{dV}{dr}$) are used as the form factors. The microscopic real and imaginary form factors have the same shape with strengths $N_{\rm R}^{\rm FF}$ and $N_{\rm I}^{\rm FF}$ respectively, where $N_{\rm R,I}^{\rm FF}$ = $N_{\rm R,I}r_{\rm
rms}^V$, where $r_{\rm rms}^V$ is the rms radius of the folded potential. In addition, form factors derived from phenomenological surface imaginary and spin-orbit potentials are included. The deformation parameters $\delta$ are determined by fitting the inelastic scattering angular distribution. Table-3 shows that the quadrupole deformations obtained from the present analysis for $^{18,20,22}$O are in excellent agreement with those extracted from B(E2) values \cite{Ra01} while that for $^{18}$Ne is significantly underestimated due to lack of experimental data at forward angles. 

\vspace{-0.2cm}
\begin{table}[hb]
\centering
\vspace{-0.5cm}
\caption{Comparison of nuclear deformation parameters $\delta$ extracted from inelastic scattering and from B(E2) values.}
\begin{tabular}{ccc}
\hline
\hline
Nucleus& $\delta$   &  $\delta$     \\ \hline
       &Present work& From B(E2) values \\
\hline
 O$^{18}$ &0.33&0.355(8)* \\ 
 O$^{20}$ &0.46&0.261(9)* [ 0.50(4)**] \\ 
 O$^{22}$ &0.26&0.208(41)* \\ 
Ne$^{18}$ &0.40&0.694(34)* \\  \hline
\hline
* From Ref.\cite{Ra01}& &** from Ref.\cite{Je99} 
\label{table3}
\vspace{-0.8cm}
\end{tabular}
\end{table}

\vspace{-0.2cm}     
\section{Nuclear decays}

\vspace{-0.2cm}
\subsection{Proton radioactivity}

    The half lives of the decays of spherical nuclei away from proton drip line by proton emissions are estimated theoretically. The half life of a parent nucleus decaying via proton emission is calculated using the WKB barrier penetration probability. The WKB method is found quite satisfactory and even better than the S-matrix method for calculating half widths of the $\alpha$ decay of superheavy elements \cite{Ma06}. For the present calculations, the zero point vibration energies used here are given by Eq.(5) of Ref.\cite{Po86} extended to protons and the experimental $Q$ values \cite{So02} are used. Spherical charge distributions are used for Coulomb interaction potentials. The nuclear potential $V_N(R)$ of Eq.(8) has been replaced by $V_N(R)+V^{Lane}_N(R)$ where the isovector \cite{Sa83} or symmetry component of the folded potential $V^{Lane}_N(R)=\int\int[\rho_{1n}(\vec{r_1})-\rho_{1p}(\vec{r_1})] [\rho_{2n}(\vec{r_2}) -\rho_{2p}(\vec{r_2})] v_{01}[|\vec{r_2} - \vec{r_1} + \vec{R}|] d^3r_1 d^3r_2$ where the subscripts 1 and 2 denote the daughter and the emitted nuclei respectively while the subscripts n and p denote neutron and proton densities respectively. With simple assumption that $\rho_{1p}=[\frac{Z_d}{A_d}]\rho$ and $\rho_{1n}=[\frac{(A_d-Z_d)}{A_d}]\rho$, and for the emitted particle being proton $\rho_{2n}(\vec{r_2})- \rho_{2p} (\vec{r_2})=-\rho_2(\vec{r_2})=-\delta(\vec{r_2})$, the Lane potential becomes $ V^{Lane}_N(R) = -[\frac{(A_d-2Z_d)}{A_d}] \int \rho (\vec{r}) v_{01} [|\vec{r} - \vec{R}|] d^3r $ where $v_{01}(s)=t_{01}^{M3Y}(s,E)g(\rho)$ and $A_d$ and $Z_d$ are, respectively, the mass number and the charge number of the daughter nucleus. The inclusion of this Lane potential causes insignificant changes in the lifetimes. The same set of data of Ref.\cite{Bal05} has been used for the present calculations using $C$=2.2497 and $\beta$=1.5934 fm$^2$. The agreement of the present calculations with a wide range of experimental data for the proton radioactivity lifetimes are reasonably good \cite{BCS08}. 

\vspace{-0.2cm}      
\subsection{Alpha radioactivity of SHE}

    The double folded nuclear potential between the daughter and emitted nuclei is given by

\vspace{-0.57cm}
\begin{equation}
 V_N(R) = \int\int \rho_1(\vec{r_1})\rho_2(\vec{r_2})v_{00}[|\vec{r_2} - \vec{r_1} + \vec{R}|]d^3r_1 d^3r_2 
\label{seqn19}
\end{equation}

\noindent
where $\rho_1$, $\rho_2$ are the density distribution functions for the two composite nuclear fragments. Since the density dependence of the effective projectile-nucleon interaction was found to be fairly independent of the projectile \cite{Sr83}, as long as the projectile-nucleus interaction was amenable to a single-folding prescription, the density dependent effects on the nucleon-nucleon interaction can be factorized into a target term times a projectile term as $g(\rho_1, \rho_2) = C (1 - \beta \rho_1^{2/3}) (1 - \beta \rho_2^{2/3})$. The parameter $\beta$ can be related to the mean free path in nuclear medium; hence its value should remain the same, 1.5934 fm$^2$, as that obtained from nuclear matter calculations, while the other constant C, which is basically an overall normalization constant, may change. The value of this overall normalization constant is kept equal to unity, which has been found $\approx$1 from an optimum fit to a large number of $\alpha$ decay lifetimes \cite{Ba03}. This formulation is used successfully in case of $\alpha$ radioactivity of nuclei \cite{Ba03} including superheavies \cite{CSB06,prc07,scb07}. In $\alpha$-decay calculations only the isoscalar term contributes because $\alpha$ contains equal number of neutrons and protons. 

\vspace{-0.2cm}    
\subsection{Cluster radioactivity}

    The decay constant $\lambda$ for cluster radioactivity is a product of cluster preformation probability $P_0$ in the ground state, the tunneling probability through barrier $P$ and the assault frequency $\nu$. The preformation factor may be considered as the overlap of the actual ground state configuration and the configuration representing the cluster coupled to the ground state of the daughter. Superheavy emitters being loosely bound than highly bound $\alpha$, $P_0$ is expected to be high for $\alpha$ decay and the present calculations with $P_0$=1 provide excellent description of $\alpha$ decay for recently discovered superheavy nuclei \cite{CSB06,prc07,scb07}. For weakly bound heavy cluster decay it is expected to be orders of magnitude less than unity. The theoretical half lives of cluster radioactivity for very heavy nuclei are calculated assuming cluster preformation factor to be unity. Hence the preformation factors $P_0$ are calculated \cite{Ro09} as the ratios of the calculated half lives to the experimentally observed half lives. 

\vspace{-0.2cm}
\section{Neutron stars}

\vspace{-0.2cm}
\subsection{Modeling neutron Stars}

    If rapidly rotating compact stars were nonaxisymmetric, they would emit gravitational waves in a very short time scale and settle down to axisymmetric configurations. Therefore, we need to solve for rotating and
axisymmetric configurations in the framework of general relativity. For the matter and the spacetime the following assumptions are made. The matter distribution and the spacetime are axisymmetric, the matter and the spacetime are in a stationary state, the matter has no meridional motions, the only motion of the matter is a circular one that is represented by the angular velocity, the angular velocity is constant as seen by a distant observer at rest and the matter can be described as a perfect fluid. The energy-momentum tensor of a perfect fluid $T^{\mu\nu}$ is given by $T^{\mu\nu}$=$(\varepsilon+P)u^\mu u^\nu-g^{\mu\nu}P$ where $\varepsilon$, $P$, $u^\mu$ and $g^{\mu\nu}$ are the energy density, pressure, four velocity and the metric tensor, respectively. To study the rotating stars the following metric is used

\vspace{-0.55cm}
\begin{equation}
ds^2 = -e^{(\gamma+\rho)} dt^2 + e^{2\alpha} (dr^2+r^2d\theta^2) + e^{(\gamma-\rho)} r^2 \sin^2\theta (d\phi-\omega dt)^2
\label{seqn22}
\end{equation}
\vspace{-0.55cm}

\noindent
where gravitational potentials $\gamma$, $\rho$, $\alpha$ and $\omega$ are functions of polar coordinates $r$, $\theta$ only. Einstein's field equations for the three potentials $\gamma$, $\rho$ and $\alpha$ are solved using the Green's-function technique \cite{Ko89} and the fourth potential $\omega$ is determined from other potentials. At zero frequency limit corresponding to the static solutions of Einstein's field equations for spheres of fluid, present formalism yields results for the solution of TOV equation \cite{TOV39}. 

\vspace{-0.2cm}
\subsection{$\beta$-equilibrated neutron star matter and quark matter EoS}

    The nuclear matter EoS,  as described earlier, is calculated \cite{BCS08} using the isoscalar and the isovector components of M3Y interaction along with density dependence which is determined completely from the nuclear matter calculations. This EoS evaluated at the isospin asymmetry $X$ determined from the $\beta$-equilibrium proton fraction $x_\beta$ [$=\frac{\rho_p}{\rho}$], obtained by solving $\hbar c (3 \pi^2\rho x_\beta)^{1/3}$=$4E_{sym}(\rho) (1 - 2 x_\beta)$, provides EoS for the $\beta$-equilibrated NS matter where $E_{sym}(\rho)$ is the NSE. For cold and dense quark (QCD) matter, the perturbative EoS \cite{Ku10} with two massless and one massive quark flavors and a running coupling constant, is used. The constant $B$ is treated as a free parameter, which allows to take into account non-perturbative effects not captured by the weak coupling expansion. 

\vspace{-0.2cm}
\subsection{Deconfinement phase transition: from nuclear matter to quark matter}

    The energy density of the quark matter is lower than that of the present EoS for the $\beta$-equilibrated charge neutral NS matter at densities higher than 0.405 fm$^{-3}$ for bag constant $B^{\frac{1}{4}}$=110 MeV \cite{Ku10} implying presence of quark core. For lower values of bag constant such as $B^{\frac{1}{4}}$=89 MeV, energy density for our EoS is lower and makes a cross over with the quark matter EoS at very high density $\sim$1.2 fm$^{-3}$ causing too little quark core (predicting similar results as NS with pure nuclear matter inside) and therefore we choose $B^{\frac{1}{4}}$=110 MeV for representative calculations. The common tangent is drawn for the energy density versus density plots where pressure is the negative intercept of the tangent to energy density versus density plot. However, the phase co-existence region is negligibly small which is represented by part of the common tangent between the points of contact on the two plots \cite{Pe93} implying constant pressure throughout the phase transition. 

\begin{figure}[htbp]
\begin{center}
\includegraphics[width=0.59\columnwidth]{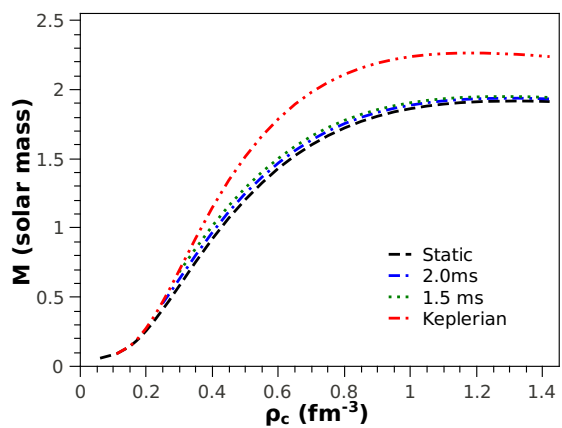}
\caption{Variation of mass with central density for static and rotating neutron stars with pure nuclear matter inside.}
\label{fig:8}
\end{center}
\end{figure}
    
\vspace{-0.2cm}    
\subsection{Calculations and results: masses and radii of neutron and hybrid Stars}

    We use the `rns' code \cite{St95} for calculating compact star properties which requires EoS in the form of energy density versus pressure along with corresponding enthalpy and baryon number density. The rotating compact star calculations are performed using the crustal EoS, FMT \cite{FMT49}+ BPS \cite{BPS71}+ BBP \cite{BBP71} upto number density of 0.0458 fm$^{-3}$ and $\beta$-equilibrated NS matter beyond. It is worthwhile to mention here that a star may not rotate as fast as Keplerian frequency due to r-mode instability. The variation of mass with central density for static and rotating neutron stars at Keplerian limit and also maximum frequencies limited by the r-mode instability with pure nuclear matter inside is shown in Fig.-2. NSs with pure nuclear matter inside, the maximum mass for the static case is 1.92 M$_\odot$ with radius $\sim$9.7 km and for the star rotating with Kepler's frequency it is 2.27 M$_\odot$ with equatorial radius $\sim$13.1 km \cite{Ch10}. However, for stars rotating with maximum frequency limited by the r-mode instability, the maximum mass turns out to be 1.95 (1.94) M$_\odot$ corresponding to rotational period of 1.5 (2.0) ms with radius about 9.9 (9.8) kilometers. When quark core is considered, the maximum mass for the static case is 1.68 M$_\odot$ with radius $\sim$10.4 km and for the star rotating with Kepler's frequency it is 2.02 M$_\odot$ with equatorial radius $\sim$14.3 km whereas stars rotating with maximum frequency limited by the r-mode instability, the maximum mass turns out to be 1.72 (1.71) M$_\odot$ corresponding to rotational period of 1.5 (2.0) ms with radius about 10.7 (10.6) kilometers \cite{APP12}. 

\vspace{-0.2cm}
\section{Summary and conclusion}

    In summary, we show that theoretical description of nuclear matter based on mean field calculation using density dependent M3Y effective NN interaction yields a value of nuclear incompressibility which is highly in agreement with that extracted from experiment and gives a value of NSE that is consistent with the empirical value extracted by fitting the droplet model to the measured atomic mass excesses and with other modern theoretical descriptions of nuclear matter. The slope $L$ and the isospin dependent part $K_\tau$ of the isobaric incompressibility are consistent with the constraints recently extracted from analyses of experimental data. We have applied our nucleonic EoS with a thin crust to solve the Einstein's field equations to determine the mass-radius relationship of neutron stars with and without quark cores. We have obtained the masses of neutron (hybrid) stars rotating with Keplerian frequencies, around 2.27 (2.02) M$_\odot$ with equatorial radii around 13 (14) kilometres. The maximum mass of NS without quark core, with maximum rotational frequency limited by the r-mode instability, turns out to be 1.95 (1.94) M$_\odot$ corresponding to rotational period of 1.5 (2.0) ms with radius about 9.9 (9.8) kilometers which is in excellent agreement with recent astrophysical observations. The nucleon-nucleon effective interaction used in the present work, which is found to provide a unified description of elastic and inelastic scattering, various radioactivities and nuclear matter properties, also provides an excellent description of the $\beta$-equilibrated NS matter which is stiff enough at high densities to reconcile with the recent observations of the massive compact stars $\sim$2 M$_\odot$ while the corresponding symmetry energy is supersoft \cite{CBS09} as preferred by the FOPI/GSI experimental data.


\vspace{-0.2cm}
\begin{thebibliography}{999}

\bibitem{Li08} B. A. Li, L. W. Chen, and C. M. Ko, Phys. Rep. {\bf 464}, 113 (2008).

\bibitem{St05} A. W. Steiner, M. Prakash, J. M. Lattimer and P. J. Ellis, Phys. Rep. {\bf 411}, 325 (2005). 

\bibitem{Da02} P. Danielewicz, R. Lacey and W.G. Lynch, Science {\bf 298}, 1592 (2002). 

\bibitem{Be77} G.Bertsch, J.Borysowicz, H.McManus, W.G.Love, Nucl. Phys. {\bf A 284}, 399 (1977).

\bibitem{Sa79} G.R. Satchler and W.G. Love, Phys. Reports {\bf 55}, 183 (1979). 

\bibitem{ke35} E. C. Kemble, Phys. Rev. {\bf 48}, 549 (1935).

\bibitem{BCS08} D. N. Basu, P. Roy Chowdhury and C. Samanta, Nucl. Phys. {\bf A 811}, 140 (2008).

\bibitem{Ba03} D. N. Basu, Phys. Lett. {\bf B 566}, 90 (2003).

\bibitem{CSB06} P. Roy Chowdhury, C. Samanta and D. N. Basu, Phys. Rev. {\bf C 73}, 014612 (2006); {\it ibid} Phys. Rev. {\bf C 77}, 044603 (2008); {\it ibid} Atomic Data and Nuclear Data Tables {\bf 94}, 781 (2008).

\bibitem{prc07} P. Roy Chowdhury, D.N. Basu and C. Samanta, Phys. Rev. {\bf C 75}, 047306 (2007).

\bibitem{scb07} C. Samanta, P. Roy Chowdhury and D.N. Basu, Nucl. Phys. {\bf A789}, 142 (2007). 

\bibitem{Ro09} T.R. Routray, Jagajjaya Nayak and D.N. Basu, Nucl. Phys. A 826, 223 (2009).

\bibitem{Gu05} D. Gupta and D. N. Basu, Nucl. Phys. {\bf A 748}, 402 (2005).

\bibitem{Gu06} D. Gupta, E. Khan and  Y. Blumenfeld, Nucl. Phys. {\bf A 773}, 230 (2006).  

\bibitem{De10} P. B. Demorest, T. Pennucci, S. M. Ransom, M. S. E. Roberts and J. W. T. Hessels, Nature {\bf 467}, 1081 (2010).

\bibitem{Sa83} A.M. Lane, Nucl. Phys. {\bf 35}, 676 (1962).

\bibitem{Sa89} C. Samanta, D. Bandyopadhyay and J.N. De, Phys. Lett. {\bf B 217}, 381 (1989). 

\bibitem{CB06} P. Roy Chowdhury and D.N. Basu, Acta Phys. Pol. {\bf B 37},1833 (2006).

\bibitem{Au03} G. Audi, A.H. Wapstra and C. Thibault, Nucl. Phys. {\bf A 729}, 337 (2003).

\bibitem {Lu03} D. Lunney, J.M. Pearson and C. Thibault, Rev. Mod. Phys. {\bf 75}, 1021 (2003).

\bibitem{Ro06} G.Royer and C.Gautier, Phys. Rev. {\bf C 73}, 067302 (2006).

\bibitem{La99} G. A. Lalazissis, S. Raman, and P. Ring, At. Data and Nucl. Data Tables {\bf 71}, 1 (1999).

\bibitem{Vr03} D. Vretenar, T. Nik\'si\'c and P. Ring, Phys. Rev. {\bf C 68}, 024310 (2003).

\bibitem{Sh09} M. M. Sharma, Nucl. Phys. {\bf A 816}, 65 (2009).

\bibitem{Kl06} T. Kl\"ahn et al., Phys. Rev. {\bf C 74}, 035802 (2006). 

\bibitem{Mu07} T. Mukhopadhyay and D.N. Basu, Nucl. Phys. {\bf A 789}, 201 (2007).

\bibitem{Da03} P. Danielewicz, Nucl. Phys. {\bf A 727}, 233 (2003). 

\bibitem{Po03} K. Pomorski and J. Dudek, Phys. Rev. {\bf C 67}, 044316 (2003). 

\bibitem{Be89} L. Bennour et al., Phys. Rev. C {\bf 40}, 2834 (1989).

\bibitem{Ak98} A. Akmal, V.R. Pandharipande and D.G. Ravenhall, Phys. Rev. {\bf C 58}, 1804 (1998).

\bibitem{Ce09} M. Centelles, X. Roca-Maza, X. Vinas and M. Warda, Phys. Rev. Lett. {\bf 102}, 122502 (2009).

\bibitem{Ch09} Lie-Wen Chen, Bao-Jun Cai, Che Ming Ko, Bao-An Li, Chun Shen and Jun Xu, Phys. Rev. {\bf C 80}, 014322 (2009).

\bibitem{Pi09} J. Piekarewicz and M. Centelles, Phys. Rev. {\bf C 79}, 054311 (2009).

\bibitem{Sa07} Hiroyuki Sagawa, Satoshi Yoshida, Guo-Mo Zeng, Jian-Zhong Gu and Xi-Zhen Zhang, Phys. Rev. {\bf C 76}, 034327 (2007).

\bibitem{Wa09} M. Warda, X. Vi\"nas, X. Roca-Maza and M. Centelles, Phys. Rev. {\bf C 80}, 024316 (2009).

\bibitem{CBS09} P. Roy Chowdhury, D. N. Basu and C. Samanta, Phys. Rev. {\bf C 80}, 011305(R) (2009); D. N. Basu, P. Roy Chowdhury and C. Samanta, Phys. Rev. {\bf C 80}, 057304  (2009).

\bibitem{Zh09} Zhigang Xiao, Bao-An Li, Lie-Wen Chen, Gao-Chan Yong and Ming Zhang, Phys. Rev. Lett. {\bf 102}, 062502 (2009).

\bibitem{We09} De-Hua Wen, Bao-An Li and Lie-Wen Chen, Phys. Rev. Lett. {\bf 103}, 211102 (2009).

\bibitem{BA82} J. Bartel, P. Quentin, M. Brack, C. Guet, H. B. Hakansson, Nucl. Phys. {\bf A 386}, 79 (1982).

\bibitem{SA97} C. Samanta, Y. Sakuragi, M. Ito, M. Fujiwara, J. Phys. G: Nucl. Part. Phys. {\bf 23}, 1697 (1997).

\bibitem{DWUCK4} P. D. Kunz, computer code DWUCK4, unpublished.

\bibitem{Ra01} S. Raman et.al., Atomic Data and Nuclear Data Tables {\bf 78}, 1 (2001).

\bibitem{Je99} J. K. Jewell, et.al., Phys. Lett. {\bf B454}, 191 (1999).

\bibitem{Ma06} S. Mahadevan, P. Prema, C.S. Shastry and Y.K. Gambhir, Phys. Rev. {\bf C 74}, 057601 (2006). 

\bibitem{Po86} D.N. Poenaru, W. Greiner, M. Ivascu, D. Mazilu, I.H. Plonski, Z. Phys. {\bf A 325}, 435 (1986).

\bibitem{So02} A. A. Sonzogni, Nucl. Data Sheets {\bf 95}, 1 (2002).

\bibitem{Bal05} M. Balasubramaniam and N. Arunachalam, Phys. Rev. {\bf C 71}, 014603 (2005).

\bibitem{Sr83} D.K. Srivastava, D.N. Basu and N.K. Ganguly, Phys. Lett. {\bf 124 B}, 6 (1983).

\bibitem{Ko89} H. Komatsu, Y. Eriguchi, I. Hachisu, Mon. Not. R. Astron. Soc. {\bf 237}, 355 (1989).

\bibitem{TOV39} R. C. Tolman, Phys. Rev. {\bf 55}, 364 (1939); J. R. Oppenheimer and G. M. Volkoff Phys. Rev. {\bf 55}, 374 (1939).

\bibitem{Ku10} A. Kurkela, P. Romatschke, A. Vuorinen, Phys. Rev. {\bf D 81}, 105021 (2010).

\bibitem{Pe93} H. Heiselberg, C. J. Pethick and E. F. Staubo, Phys. Rev. Lett. {\bf 70}, 1355 (1993).

\bibitem{St95} N. Stergioulas, J. L. Friedman, Astrophys. J. {\bf 444}, 306 (1995).

\bibitem{FMT49} R. P. Feynman, N. Metropolis and E. Teller, Phys. Rev. {\bf 75}, 1561 (1949).

\bibitem{BPS71} G. Baym, C. J. Pethick and P. Sutherland, Astrophys. J. {\bf 170}, 299 (1971).

\bibitem{BBP71} G. Baym, H. A. Bethe and C. J. Pethick, Nucl. Phys. {\bf A 175}, 225 (1971).

\bibitem{Ch10} P. R. Chowdhury, A. Bhattacharyya and D. N. Basu, Phys. Rev. {\bf C 81}, 062801(R) (2010).

\bibitem{APP12} Abhishek Mishra, P. R. Chowdhury and  D. N. Basu, Astropart. Phys. {\bf 36}, 42 (2012).

\end{thebibliography}
\end{document}